\documentclass[twocolumn, prb, showpacs, superscriptaddress]{revtex4-1}
\pdfoutput=1
\usepackage[utf8x]{inputenc}
\usepackage{ucs}
\usepackage{amsmath}
\usepackage{amsfonts}
\usepackage{amssymb}
\usepackage{makeidx}
\usepackage{cellspace,booktabs}
\usepackage{natbib}
\usepackage{lipsum}
\usepackage{bm}
\usepackage{bbm}
\usepackage{relsize}
\usepackage{float}
\usepackage{wrapfig}
\usepackage{multirow}
\usepackage{ulem}
\usepackage{xcolor}

\usepackage{hyperref}
\hypersetup{colorlinks = true, linkcolor=magenta,citecolor=blue, urlcolor=magenta, bookmarksnumbered =  true}

\usepackage{color}
\definecolor{light-gray}{gray}{0.55}

\usepackage{microtype}

\usepackage{graphicx}

\begin{document}

\clearpage

\begin{abstract}
Recent exploration of the commensurate structure in the turbostratic double layer graphene shows that the large angle twisting can be treated by the decrease of the effective velocity within the energy spectra of the single layer graphene. Within our work, we use this result as a starting point, aiming towards understanding the physics of by a large angle twisted double layer graphene (i.e. Moire) quantum dot systems. We show that within this simple approach using the language of the first quantization, yet another so far unrealized (not up to our knowledge), illustrative property of the commutation relation appears in the graphene physics. Intriguingly, large twisting angles show to be a suitable tunning knob of the position symmetry in the graphene systems. Complete overview of the large angle twisting on the considered dot systems is provided.
\end{abstract}

\date{\today}
\author{Jozef Bucko}
\affiliation{Institute for Theoretical Physics, ETH Zurich, CH-8093, Switzerland}
\affiliation{Institute for Computational Science, University of Zurich, Winterthurerstrasse 190, 8057 Zurich, Switzerland}

\author{Franti\v{s}ek Herman}
\affiliation{Institute for Theoretical Physics, ETH Zurich, CH-8093, Switzerland}
\affiliation{Department of Experimental Physics, Comenius University, Mlynská Dolina F2, 842 48 Bratislava, Slovakia}
\title{Large twisting angles in Bilayer graphene Moire quantum dot structures}

\maketitle


\section{Introduction}

Graphene, elegant honeycomb structured atomic monolayer material, has already shown to have interesting mechanical \cite{Papageorgiou_2017} as well as electronic \cite{Neto_2009} properties of the ideal semi-metal. Due to its linear dispersion, related mathematical description as well as resulting physical properties, understanding and describing graphene unquestionably belongs to the showcase of the twenty-first century physics.\\

Adding another graphene layer shows to be an important step towards the simple way of creating the gap in the energy spectrum \cite{Min_2007}. This natural progress in the development of graphene systems also opens the field of twistronics by introducing another degree of freedom through the possibility of twisting the layers towards each other \cite{Carr_2017, Cao_2018}. New macroscopically tunable length scale emerges, resulting from the appearing structure of the Moire pattern. Its presence leads to lowering of the slope of the energy band close to the valley\cite{Moon_2012} explained by the reduction of the Fermi velocity\cite{Shallcross_2010}. Also, additional Dirac points as well as flattening band at the magic angle $\theta^{*}\approx 1.1^{\circ}$ \cite{MacDonald_2011} occur. Described tuning mechanism shows to lead towards interesting and novel properties of the underlying density of states in the normal state \cite{Carr_2017}, as well as to influencing e.g. the value of $T_c$ in the superconductive state \cite{Cao_2018}.\\

Area of twistronics becomes also attractive from the point of view of dot and flake (mesoscopic) systems bounded by the additional confining potential, which keeps the electron confined on the suitable length scale under $\sim 100\, nm$. \emph{Dot twistronics} might be even more attractive from the application point of view, since we do not run into the dimension vs. quality of the prepared systems kind of problems\cite{Greplova_2020}. Simple exploration of the energy scale of the created bound states (theoretically and experimentally), their behavior with changing characteristic length, twisting angle, as well as application of the external magnetic field create standard tools in the developing area of the dot twistronics \cite{Tiutiunnyka_2019, Mirzakhani_2020}.\\

In what follows, we introduce reduced velocity in the connection with the single layer graphene quantum dot model, in order to simulate the effect of twisting layers of the dot about large angles. In the next sections, we will have a look at the effect of twisting angles on the scaling of the states, bending of the energy levels and effect of the large twisting angles on the wave-function of the twisted bilayer graphene Moire quantum dot (BGM-QD) system.

\section{(Large angle) Twisted bilayer graphene quantum dots}\label{sec:BGM-QD}

\begin{figure}[htpb]
    \centering
    \includegraphics[scale=0.17]{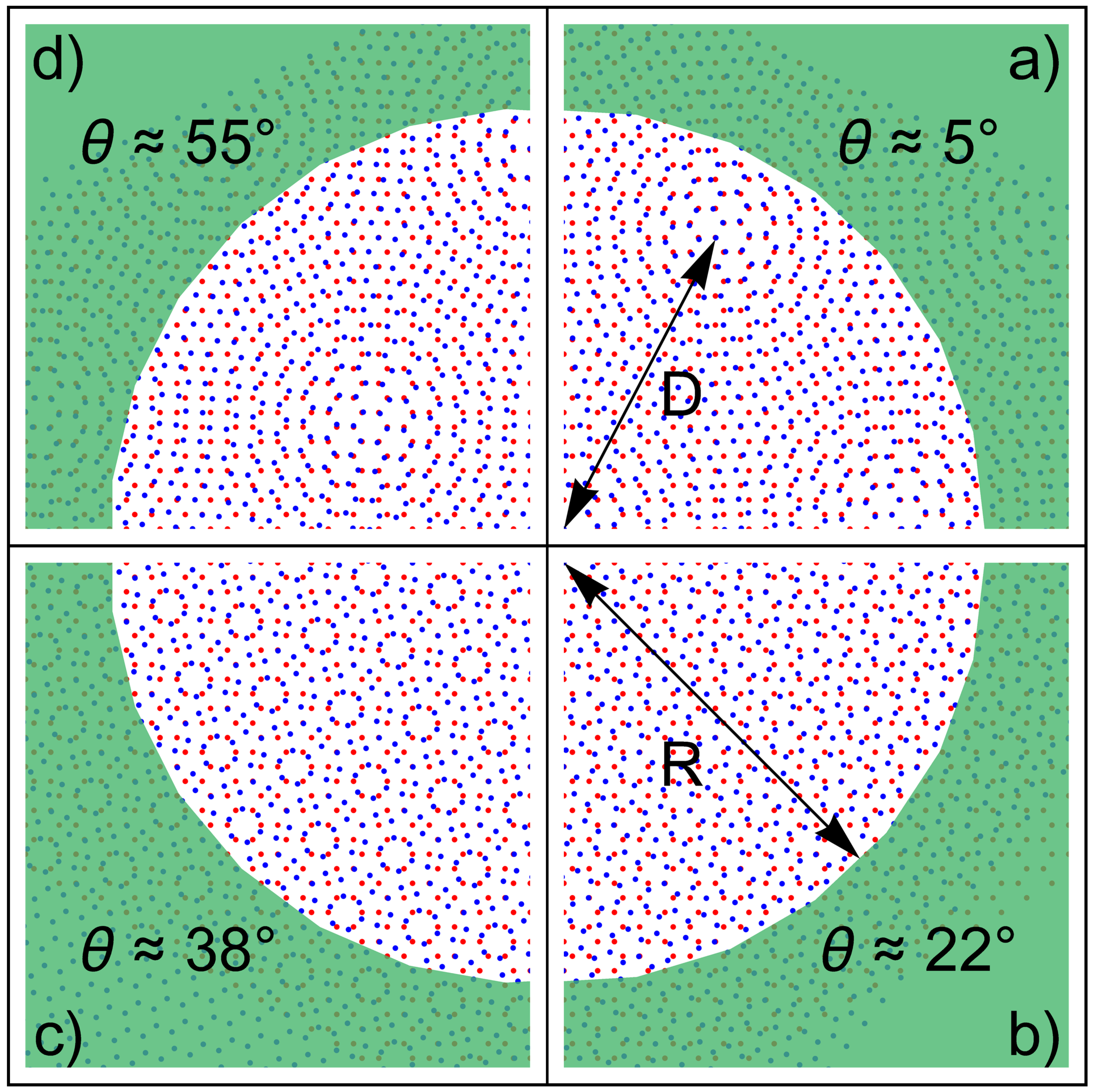}
    \caption{Moire pattern situation close to the boundary of the cartoon $7\,nm$ wide BGM-QD system considering different values of the large twisting angle $\theta$.}
    \label{fig:Moire_boundary}
\end{figure}

In the following sections, we focus on the semianalytic approach towards the explanation of the generic behavior of the energy bands of the twisted bilayer graphene quantum dot, considering large twisting angles $\theta$. Recent numerical studies published in Ref.~\onlinecite{Tiutiunnyka_2019, Mirzakhani_2020} are based on tight-binding approximation and suited as an enlightening motivation of our work. Mentioned purely numerical approach has its limitations in computational power and therefore usually allows one to focus on dot-systems considering radiuses up to $\sim\, 10\, nm$.\\

Our semianalytical approach, which allows us to examine general physical phenomena considering larger systems, is motivated by the results of the second mentioned study \cite{Mirzakhani_2020}. Its authors claim that considering the interval of large twisting angles $10^{\circ} \lesssim \theta \lesssim 50^{\circ}$, interaction between the two single sheets of graphene layers is small and the band structure basically copies the one of the two single layer sheets.\\

In the Fig.~\ref{fig:Moire_boundary}, we plot the Moire pattern structure for few chosen values of $\theta$, together with two different important length scales considered in our situation. $R$ marks the radius of our dot, while $D$ corresponds to the Moire length scale. From now on, we properly define the term BGM-QD (bilayer graphene Moire quantum dot) as regime when $R \gtrsim D$ (i.e. when the Moire pattern is clearly recognizable in the region of the dot).\\

\subsection{Effective velocity}

\begin{figure}[htpb]
\centering
\includegraphics[scale=0.32]{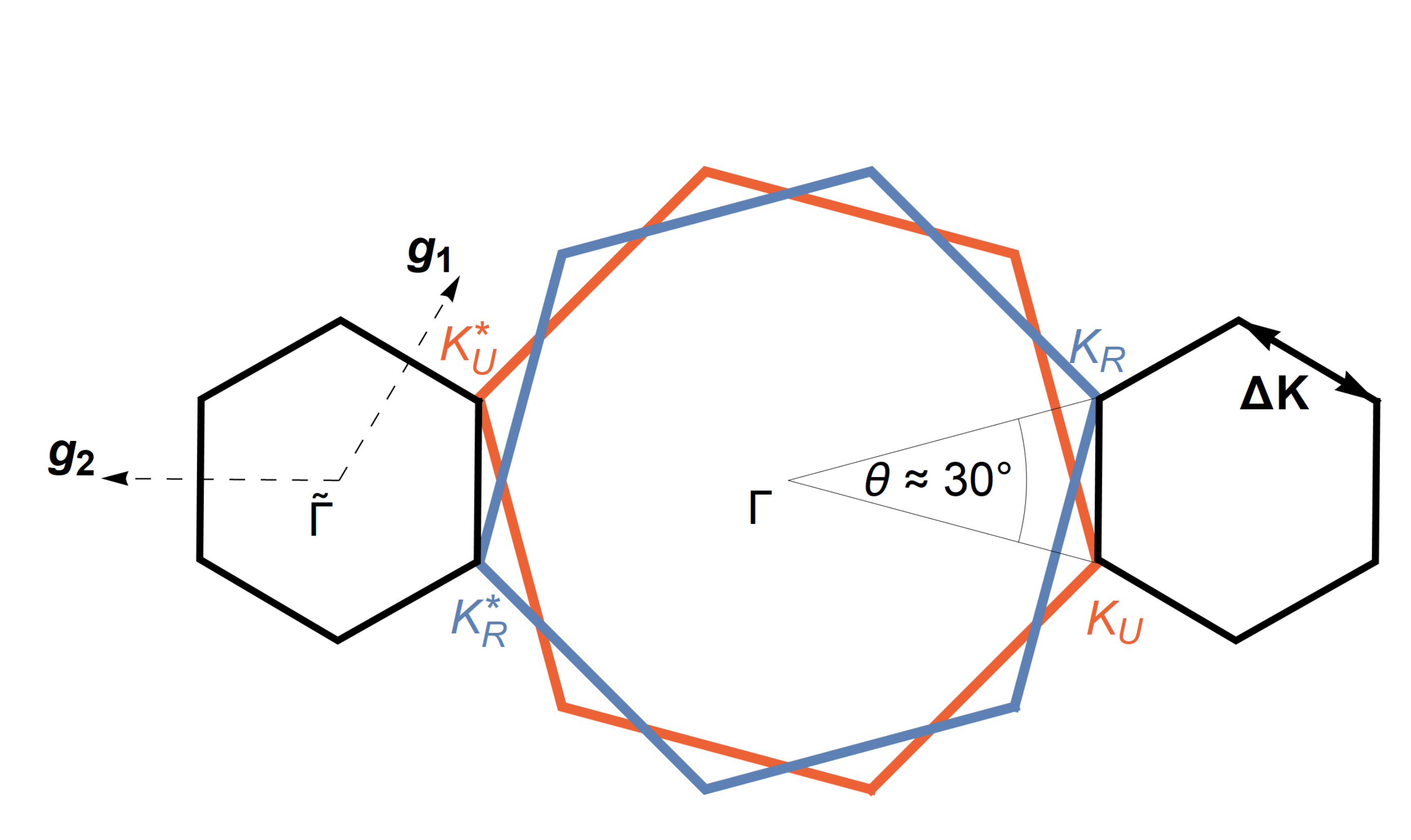}
\caption{Brillouin zone of the unrotated (red), rotated (blue) single layer graphene, as well as twisted bilayer graphene (black).}
\label{fig:BZ_Moire}
\end{figure}

In fact, our simple approach combines the single layer quantum dot model solved in the Ref.~\onlinecite{Recher_2009} together with the effective Fermi velocity close to the minimum of each valley (already measured in the Ref.~\onlinecite{Moon_2012}) within the second order of the perturbation theory suited to the crystal and electronic structure of the turbostratic graphene \cite{Shallcross_2010}:
\begin{equation}\label{eq:v_F}
    v_F = v_{F}^{SL}\left(1-\frac{\alpha}{\Delta K^2}\right),
\end{equation}
where $\alpha$ represents the binding constant element between states in individual graphene layers. Single layer Fermi velocity is represented by $v_{F}^{SL}$. The distance of valleys from the individual graphene layers $\Delta K$, displayed in the Fig.~\ref{fig:BZ_Moire}, is in the Moire structures tunable variable and depends on the twisting angle $\theta$ by \cite{Shallcross_2010}:
\begin{equation}\label{eq:K}
    \Delta K = \frac{4}{3 a_0}\sin{\frac{\theta}{2}},
\end{equation}
where $a_0 = 0.142\, nm$ is the shortest distance between the Carbon atoms in the graphene.\\

Naturally, since the Moire period $D$ can be easily connected together with the twisting angle $\theta$ according to \cite{Shallcross_2010}:
\begin{equation}\label{eq:D}
    D = \frac{\sqrt{3}}{2}a_0\csc{\frac{\theta}{2}},
\end{equation}
$\Delta K$ relates with the Moire period by simple equation $\Delta K.D = 2/\sqrt{3}$. For illustration, we plot the standard picture of two mutually rotated Brillouin zones considering two graphene sheets in the Fig.~\ref{fig:BZ_Moire}. One (light red) corresponds to the unrotated BZ of the graphene layer, while the other (light blue) corresponds to the rotated one. The picture describes the situation corresponding to the large twisting angles $\theta \approx 30^{\circ}$. In the Fig.~\ref{fig:BZ_Moire}, we also plot the individual valleys of the unrotated $\left(\mathbf{K}_U, \mathbf{K}^*_U \right)$ as well as the rotated layer $\left(\mathbf{K}_R, \mathbf{K}^*_R \right)$ in the Brillouin zone corresponding to the twisted bilayer graphene (and therefore also BGM-QD) according to the mapping described in the Ref.~\onlinecite{Shallcross_2010}.\\

Within our treatment, we consider interval of twisting angles $5^{\circ} \lesssim \theta \lesssim 55^{\circ}$, where the approximation for the effective reduced velocity in the form of the Eq.~\eqref{eq:v_F} holds. This assumption is also reasonable, once we consider our condition $R \gtrsim D$, together with the Eq.~\eqref{eq:D} and $R_{min} = 10\,nm$. After the simple calculation we find the minimal value of the twisting angle $\theta_{min} = 1.4^{\circ}$, which is way bellow our analytical approximation of the reduced velocity. Larger values of the radius $R$ lead to even smaller values of the $\theta_{min}$. Notice, that considering cases of the large twisting angles $\theta \approx 30^{\circ}$, $\Delta K$ is close to its maximum and the correction in the reduced Fermi velocity corresponding to the second order of the perturbation theory is therefore fully justified.\\

On the contrary, considering small twisting angles, the system starts to be dominated by the flat dispersion behavior coming from the merging of the two Van Hove singularities \cite{Peres_2019, MacDonald_2011}, which makes it unsuitable for our description by the model with the linear dispersion. Also, it is good to emphasize that the idea of using effective velocity in the form of the Eq.~\eqref{eq:v_Fef} allows us (in the approximation of large $\theta$) to solve two problems (two layers) at once. Therefore, our solution for the energy bands will automatically have $2 \times 2$ degeneracy assuming layers $\times$ spin.

\subsection{Hamiltonian}

After the proper introduction of the efective velocity $v_F(\theta)$ from the Eq.~\eqref{eq:v_F}, now it is time to introduce it to our effective Hamiltonian \cite{Beenakker_2008} valid close to the Dirac points. Since the energy scale that we are focusing on is much smaller than the Fermi energy for graphene, we can allow ourselves to work within the assumption of the linearized tight binding approximations close to the Dirac points. The resulting differential equation of motion is then coming from the usage of the first quantized language, where the momentum coordinates are exchanged by the differential operators of derivatives in position. Next, we will also consider rotational symmetry of the confining potential $U(\boldsymbol{r}) = 0$, assuming $r \leq R$ resp. $U(\boldsymbol{r}) = U_0$, assuming $r > R$:
\begin{equation}\label{eq:Ham}
    H^{\tau}(\theta) = v_F(\theta)\left(\mathbf{p}+e\mathbf{A}\right).\boldsymbol{\sigma} + \tau\Delta\sigma_{z} + U\left(\boldsymbol{r}\right),
\end{equation}
where $\mathbf{p}$ is the momentum operator, $\boldsymbol{\sigma} = \left(\sigma_x, \sigma_y\right)$ is the vector of the Pauli matrices and $\mathbf{A}$ is the vector potential related to the magnetic field in the $z$-direction by $\mathbf{A} = B/2\left(-y,x,0\right)$. Second term corresponds to the constant mass term creating gap $2\Delta$ induced by the underlying substrate \cite{Giovannetti_2007, Zhou_2007}, where $\tau$ differentiates the valleys.\\

To avoid any confusion, let us discuss monolayer structure of the Hamiltonian in the Eq.~\eqref{eq:Ham}. As we have already mentioned, in the regime of large twisting angles, we are in the situation, where the limit of the BGM-QD being represented as two single layer sheets of graphene as well as the introduced treatment of the reduced velocity are valid. Therefore, the Eq.~\eqref{eq:Ham} models twice the same circular monolayer graphene  flake structure (corresponding to the twisted bilayer) and thus, effectively, we solve the same problem twice (two layers, two Hamiltonians, two sets of Dirac equations). In principle, one can also solve the problem as set of two Dirac equations (each corresponding to separate layer), while considering two different gap terms. First suited to one layer and the second to another layer. The difference would be at the end in the splitting of the degeneracy of the states coming from two layers in our case. Otherwise, the qualitative features of the provided solution would be the same as the ones described in the next chapters.\\

Therefore, our model corresponds to the (theoretically) simplest realization where we assume using the substrate for one graphene flake (applied from one side of the graphene layer) and then the same substrate from the other side of the second graphene layer. Our two twisted graphene flakes would be therefore sandwiched by the same substrate, which would create the same gap in both of them. It is fair to say, that even though this model represents the simplest theoretical realization, it may be experimentally challenging.\\

Important thing to mention and also make clear is the difference between the geometry of the gap and bounding potential of the dot in our model compared to already discussed numerical tight-binding studies. In our approach (which is close to the one used in the Ref.~\onlinecite{Recher_2009}) the gap is realized in the whole region of the dot and the confining potential $U(\boldsymbol{r})$ is nonzero only for $r > R$. However in the numerical study in Ref. \onlinecite{Mirzakhani_2020}, the dot is being created by the gating of the region outside of the dot.

\subsection{Uncertainty principle}\label{subsec:Up}

In the following, let us consider following commutator of the Hamiltonian \eqref{eq:Ham} (considering value of the vector potential $\mathbf{A} = \mathbf{0}$) together with the position operator $\mathbf{r}$:
\begin{align}\label{eq:commutator}
    \left[H^{\tau}(\theta),\mathbf{r}\right] &= v_F(\theta)\left[\mathbf{p},\mathbf{r}\right]\cdot\boldsymbol{\sigma},\nonumber\\
                                             &= -i\hbar v_F(\theta)\boldsymbol{\sigma},
\end{align}
where the well known canonical commutation relation $\left[p_a,r_b\right] = -i \hbar \delta_{ab} $ was used. In general, results of the commutators are interesting in two cases. First, once they are zero and we start to think about common eigenstates of the considered operators (together with their corresponding eigenvalues), symmetries and conservation laws (if one of the operators is Hamiltonian of the system). Or second, once the commutator is constant and therefore we have uncertainty principle available.\\

Notice, that the Eq.~\eqref{eq:commutator} offers a compromise of these two properties. Notice also, that the twisting angle $\theta$ entering the reduced effective velocity works as a tuning knob of the deviance from the exact symmetry of the position operator of the Hamiltonian. In such a way, due to the unique properties of the graphene Hamiltonian, we have non-trivial commutation relation where the macroscopically accessible parameter tunes the quantum properties of the system. If (assuming theoretical limit) $v_F(\theta)=0$, then the mentioned operators have common eigenstates with their corresponding eigenvalues. One would expect that the rising constant on the right side of the Eq.~\eqref{eq:commutator} will measure deviance from this case. Since the Hamiltonian will still have an exact energy eigenvalue due to Dirac equation, the state will be no longer precisely set in position. In simpler words, Eq.~\eqref{eq:commutator} means that the slower Dirac electron is the one which is more localized.\\ 

If we think about the overall effect of the twisting angle $\theta$ in the bilayer graphene, we can easily qualitatively sum it up by saying, that the large angle (macroscopically tunable) twist corresponds to the creating additional potential rising from the new periodic length scale $D$. Treating this additional potential on the level of the second order perturbation theory assuming two sheets of graphene shows, that the overall result leads towards slowing down (reduction of the velocity) of the Dirac electron once we assume decreasing twisting angles from the value of $\theta=30^{\circ}$.\\

On the top of this interesting property, we can still explore underlying uncertainty principle related to the Eq.~\eqref{eq:commutator}. Let us remind the Schrödinger uncertainty relation \cite{Griffiths_2005} suitable for our situation:
\begin{equation}
    \Delta E \Delta \mathbf{r} \geq \sqrt{\left(\frac{1}{2}\langle\lbrace H,\mathbf{r} \rbrace\rangle - \langle H \rangle\langle \mathbf{r} \rangle\right)^2 + \left(\frac{1}{2i}\langle\left[ H,\mathbf{r}\right]\rangle\right)^2},
\end{equation}
where $\Delta E^2 = \langle\psi \left(H - E\right)|\left(H - E\right)\psi\rangle$ and $\Delta \mathbf{r}^2 = \langle\psi \left(\mathbf{r} - \langle \mathbf{r} \rangle\right)|\left(\mathbf{r} - \langle \mathbf{r} \rangle\right)\psi\rangle$ . Now, thanks to the Hamiltonian being one of the operators, the equation of motion (Dirac equation) $H|\psi\rangle=E|\psi\rangle$ and also the fact that in our considered system we assume radial symmetry $\langle \mathbf{r} \rangle = 0$, we get:
\begin{equation}\label{eq:comm_E_r}
    \boxed{\Delta E \Delta \mathbf{r} \geq \mathbf{0}.}
\end{equation}
Notice, that this uncertainty principle allows us in principle to have well localized states together with the exact values of energy. All that due to the symmetry of our system and the equation of motion.\\

To be more general and also a bit pedagogical, Eq.~\eqref{eq:comm_E_r} is exactly the reason why all of the particles in the boxes (where $\langle \mathbf{r} \rangle = 0$) etc. are so grateful objects of our focus in quantum mechanics. All of these systems allow us to have discrete energy levels, together with the localized states on the reasonable length scales.\\

\subsection{Reduced energy scale}\label{subsec:redEnScale}

However enlightening we can find the Eq.~\eqref{eq:comm_E_r} to be in the strictly mathematical sense, in the real life experimental conditions we would expect $\Delta \mathbf{r}$ to be on the same scale as the radius of the dot $R$. Together with the Eq.~\eqref{eq:commutator}, we can get an estimation for the energy level resolution:
\begin{equation}\label{eq:Heinsenberg}
    \Delta E R \approx \hbar v_F(\theta).
\end{equation}
This equation makes complete sense together with the Eq.~\eqref{eq:commutator}, under which we would expect that once the velocity reduces, the system will go closer to the classical one, with more dense energy levels (closer to the continuum since the position operator is already continuous).\\

So, the other idea of this part can be summarized by the statement that the reduced Fermi velocity, which leads to the reduction of the energy scale occurring in the graphene physics:
\begin{equation}\label{eq:v_Fef}
    \frac{\hbar v_F^{SL}}{R} \rightarrow \frac{\hbar v_F(\theta)}{R},
\end{equation}
affects the behavior of the allowed energies as well as states in the dot problem with the twisting angle $\theta$.\\

\section{Formulation of the eigenproblem}

After the short discussion focused on the physics of the considered system, based on the analysis of the uncertainty principle, let us now introduce equations of motion together with the corresponding boundary condition. Mesoscopic system of the BGM-QD considering large twisting angles is on the scale of $\sim 10\, nm$ in space and $\sim 10\, meV$ in energies.\\

Based on our previous discussion related to the Hamiltonian defined by the Eq.~\eqref{eq:Ham}, Dirac equation of motion describing our BGM-QD system reads:
\begin{equation}\label{eq:H0b_1}
H^{\tau}(\theta) \Psi^{\tau}(\theta) = E^{\tau}(\theta) \Psi^{\tau}(\theta),
\end{equation}
where the Brillouin zone of the system has two sites and we order the components of the two-site envelope wave-function in the following way\cite{Recher_2009}:
\begin{equation}
\Psi^{\tau}(\theta) = 
    \begin{pmatrix}
        \Psi_{A}^{\tau}(\theta)\\
        \Psi_{B}^{\tau}(\theta)
    \end{pmatrix}.
\end{equation}
Let us remind, that we explicitly wrote down the dependence on the new external parameter of the twisting angle $\theta$.
The two-site envelope wave function (due to cylindrical symmetry) can be factorized as:
\begin{equation}
\label{eq:transform1}
\Psi^{\tau}(r,\varphi) = 
\frac{e^{ i m \varphi}}{\sqrt{r}}
\begin{pmatrix}
1 & 0 \\
0& e^{- i \varphi} \\
\end{pmatrix}
\Psi^{\tau}_1(r),
\end{equation}
where $r$ and $\varphi$ are the polar spatial coordinates and $\Psi^{\tau}_1(r)$ is the remaining part of the envelope wave-function. In the Appendix \ref{sec:Appendix_A_Sol_Strategy}, we provide further explanation and simplification of the introduced eigenvalue problem described by the Eq.~\eqref{eq:H0b_1} up to the complete recipe of the half analytical (angular part) and half numerical (radial part) solution.\\

After the introduction of the considered eigenvalue problem, let us emphasize, that the final eigenenergies as well as final form of the allowed eigenvectors can be found from the solution of the boundary condition (at $r = R$) using states inside $(<)$ and outside $(>)$ of the dot:
\begin{equation}\label{eq:Boundary}
    \Psi^{\tau}_{<}(R, \varphi) = \Psi^{\tau}_{>}(R, \varphi).
\end{equation}
The difference between solutions inside and outside of the dot is of course coming from the bounding potential $U(\boldsymbol{r})$ defined by the Eq.~\eqref{eq:Ham}, which is zero (nonzero) inside (ouside of the dot). 

\section{Results and discussion}

After introducing realized model as well as describing its solution, we can finally address the effect of the reduced effective velocity on the considered energy levels, their behavior in the magnetic field and also their angular dependence. At the end, we look at the behavior of the considered wave function evolving together with the twisting angle $\theta$. For clarity, we will focus only on the case considering $m=0$.

\subsection{State scaling}

\begin{figure}[htpb]
\centering
\includegraphics[scale=0.18]{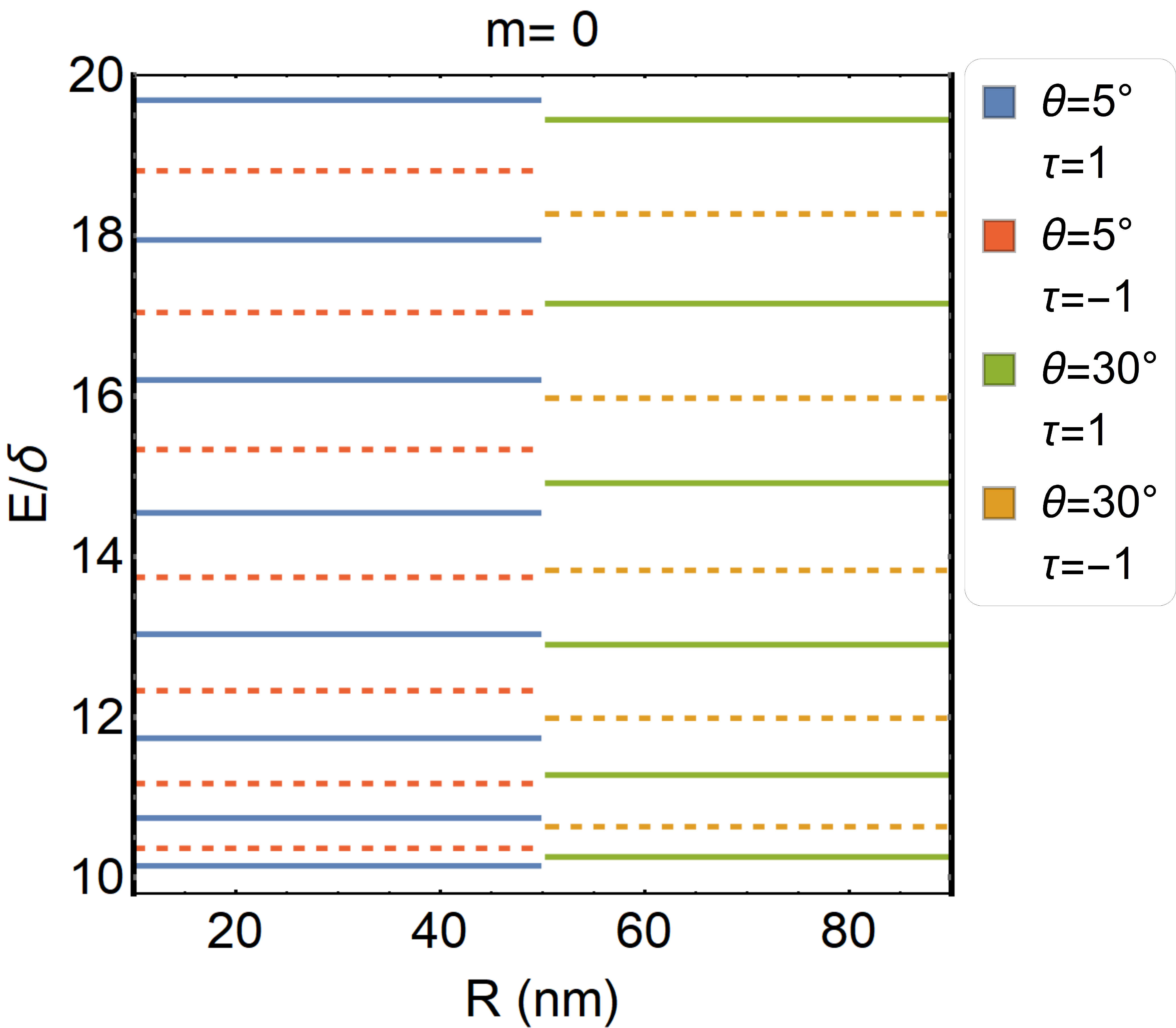}
\caption{Energy levels of the BGM-QD considering $m=0$, potentials $U_0 = \Delta = 10\delta$, where $\delta = \hbar v_F(\theta)/R$, two different valleys $\tau = \pm 1$ ($+1$ solid, $-1$ dashed), as well as two different values of twisting angle $\theta = 5^{\circ}$ (red and blue) and $\theta = 30^{\circ}$ (green and orange).}
\label{fig:R_scaling}
\end{figure}

In the Fig.~\ref{fig:R_scaling}, we plot the $R$-scaling of the states considering two different values of the twisting angle $\theta$. Notice, that the twisting angle $\theta$ does not change the $R$-evolution of the states, but works as a tuning knob of the distance between the energy levels. Provided numerical solution therefore agrees nicely with our understanding gained from the uncertainty principle formulated in the Sec.~\ref{subsec:Up}.

In the very rough approach the energy levels can be approximated by the formula: $E = \hbar v_F(\theta) g\left(m,\tau\right)/R$ (we will see that the situation is slightly more complicated). Function $g\left(m, \tau \right)$ is discrete function of $m$ as well as $\tau$ obeying all symmetries already discussed in the Ref.~\onlinecite{Recher_2009} and energy states have clear $1/R$ scaling. Notice also, that in the limit of the small radiuses, our rather general semi-analytical analysis qualitatively agrees and also generalizes the one realized and numerically calculated in the Ref.~\onlinecite{Mirzakhani_2020}.

\subsection{Effect of the magnetic field on the states and energy levels}

\begin{figure}[htpb]
\centering
\includegraphics[scale=0.18]{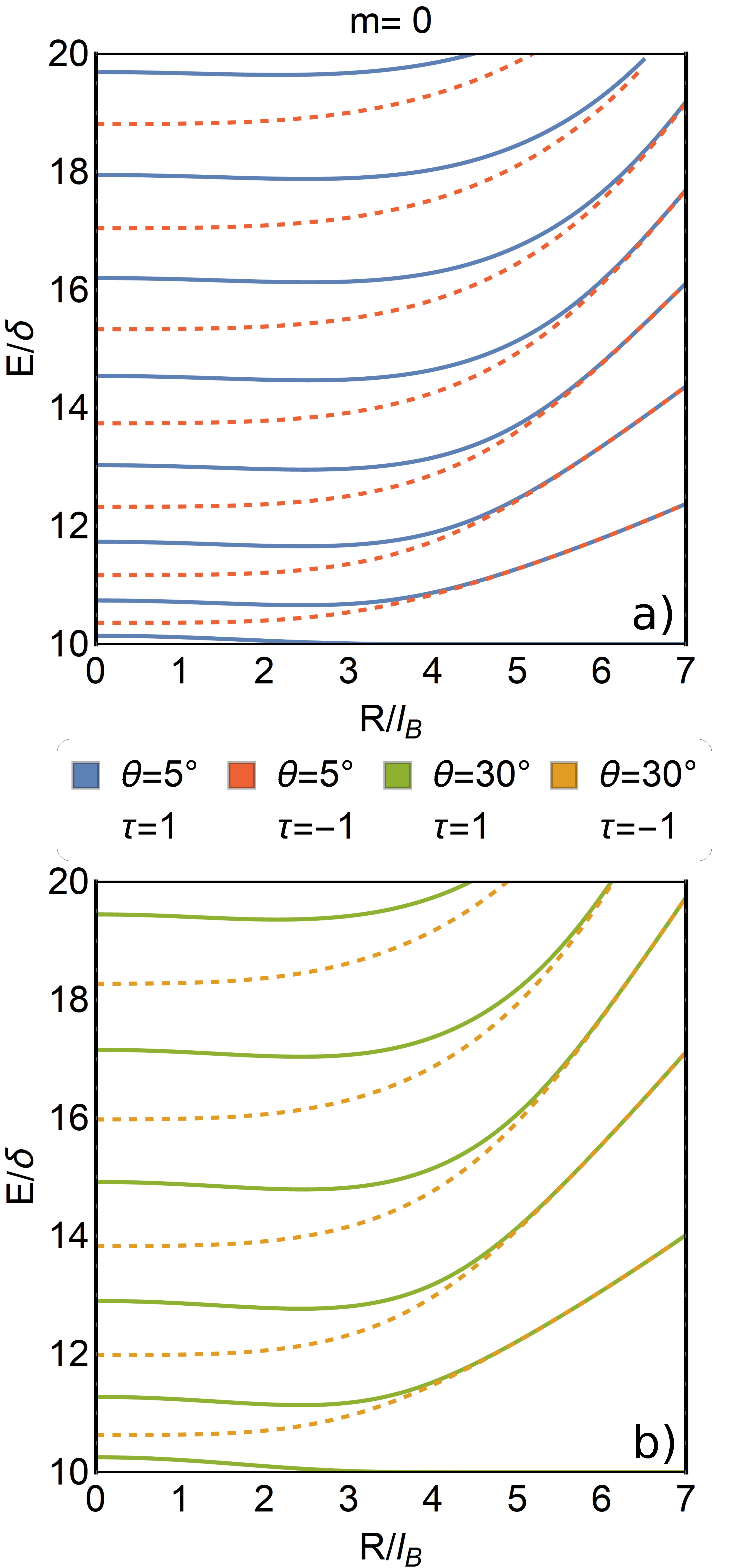}
\caption{Energy levels in the magnetic field considering twisting angles a) $\theta = 5^{\circ}$ and b) $\theta = 30^{\circ}$.}
\label{fig:mag_f_landau_lvls}
\end{figure}

Let us start by the Fig.~\ref{fig:mag_f_landau_lvls}, where we compare evolution of the energy levels in the magnetic field considering two different values of the twisting angles  $\theta = 5^{\circ}$ and $\theta = 30^{\circ}$. In the overall paragraph, we use dimensionless suitable units of the fraction $R/l_B$, where $l_B = \sqrt{\hbar/e B}$. For clarity, we focus just on the states with $m = 0$ and $\tau = \pm 1$. As we can clearly see, the main difference results from the already discussed scaling of the states with the twisting angle $\theta$. This effect then represents itself in the smaller distances between corresponding energy levels assuming lower values of the twisting angle $\theta$.

\subsubsection{Wave-function}

\begin{figure}[htpb]
\centering
\includegraphics[scale = 0.2]{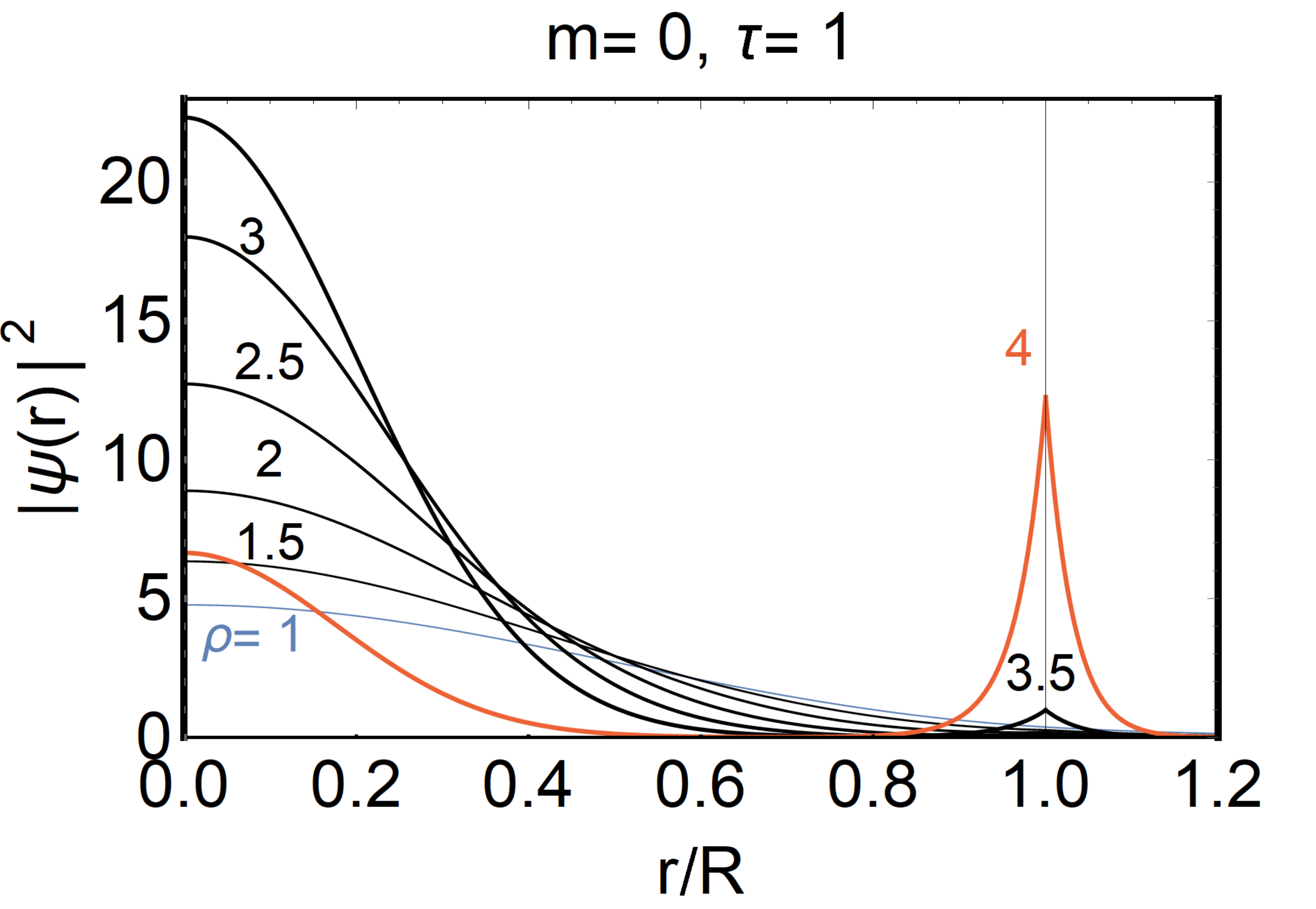}
\caption{Wave function of the twisted bilayer quantum dot as a function of the increasing ratio of $R$ and magnetic length $l_B$: $\rho = R/\sqrt{2}l_B$. Parameters related to the considered model: $m = 0$, $\tau = 1$, $U_0 = 2\delta$, $\Delta = 2\delta$, $R = 25\, nm$, $\theta = 30^{\circ}$.}
\label{fig:Wave_fun_rho}
\end{figure}

Let us now focus on the effect of the magnetic field on the wave-function of the underlying state. In the Fig.~\ref{fig:Wave_fun_rho}, we plot the BGM-QD wave-function at different values of $\rho = R/\sqrt{2}l_B$. These units are motivated by the idea of counting the allowed states $\mathcal{N}$ inside the dot, since according to Ref.~\onlinecite{Tong_2016} $\mathcal{N} = R^2/2l_B^2$. We can immediately notice, that considering rising $\rho$ the state becomes more localized in the centre. This property is in complete agreement with the logic that the increasing magnetic field allows us to have a better localized state inside the dot with the radius $R$. We know, that this is true due to the uncertainty principle based on the commutation relation between coordinates of the orbit in the magnetic field with the center at coordinates $(X,Y)$ \cite{Tong_2016}:
\begin{equation}
\left[X,Y\right] = 2 i l_{B}^{2}, \quad \quad \quad \Delta X \Delta Y = l_{B}^{2}.
\end{equation}
I.e. the weight of the wave-function lies therefore more and more inside the dot. Considering increasing value of $\rho$ going towards the limit $\rho \gg 1$, the creation of the Landau levels (plotted already in the Fig.~\ref{fig:mag_f_landau_lvls}) allows us to have another area of higher probability density localized around the boundary of the dot.\\

Fig.~\ref{fig:Wave_fun_rho} also opens discussion in the direction of the applicability of the graphene quantum dots as promising candidate for the quantum bit material. It shows, that in the quantum dots being created by the gates where $\Delta$ (or voltage between layers $V$, if we think about double layer graphene system \cite{Recher_2009}) is on the same scale as the bounding potential $U$ \cite{Eich_2018}, huge part of the wave-function is being localized outside of the dot. As we can see, applying perpendicular magnetic field (considering right amplitude) can lead to better localization of the particle inside the dot itself and therefore to more stable states with higher decoherence times.

\subsection{Effect of the twisting angle on the states and energy levels}

\begin{figure}[htpb]
\centering
\includegraphics[scale=0.17]{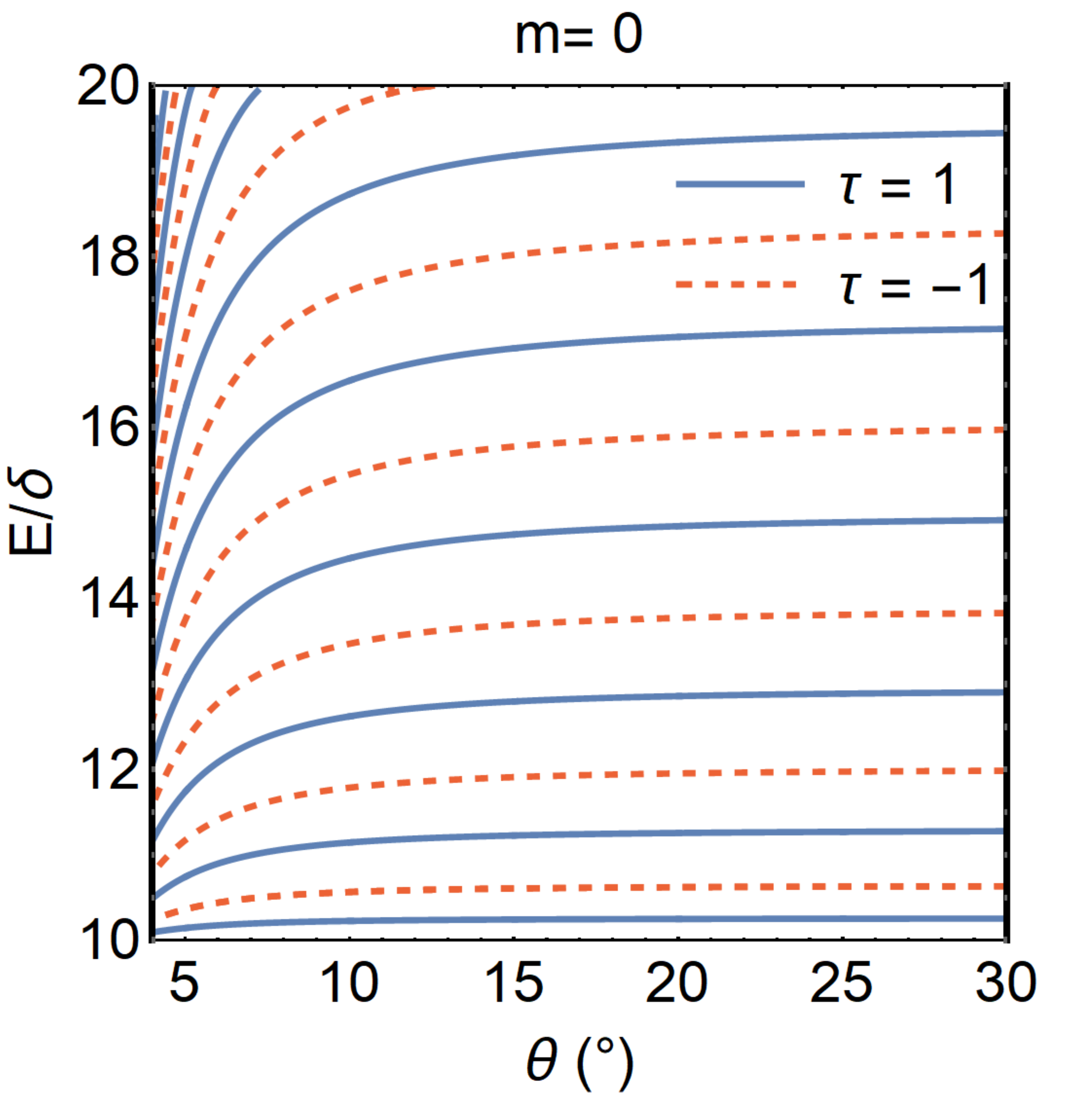}
\caption{Evolution of the states in the conduction band as a function of the twisting angle $\theta$ considering both valleys $\tau = \pm 1$.}
\label{fig:Angle_dep}
\end{figure}

In the Fig.~\ref{fig:Angle_dep} we plot the energy levels as a function of the twisting angle $\theta$. Notice, that qualitatively we obtain expected behavior of decreasing energy scale towards smaller angles (bending bands). One can easily notice, that the decreasing trend can not be explained purely by the decrease of the effective velocity $v_F(\theta)$ from the Eq.~\eqref{eq:v_F} in the form described by the Eq.~\eqref{eq:v_Fef}.
However, each of the curves plotted in the Fig.~\ref{fig:Angle_dep} can be in fact easily fitted by an empirical formula:
\begin{equation}
    \varepsilon' = \varepsilon_0 \left( 1-\gamma \csc^2{\left(\frac{\theta}{2}\right)} \right),
\end{equation}
considering fitting parameters $\varepsilon_0$ and $\gamma$. This formula should not be in fact surprising, since it reminds us the combination of the mentioned formulae Eq.~\eqref{eq:v_F}, \eqref{eq:K} (which are valid up to second order of the perturbation theory), together with the free binding parameter between the states in the separate layers. Since the overlap between different states and Hamiltonian correction will be different, it makes sense that $\gamma$ will differ. Notice also that the twisting in our considered system respects the symmetry of the system, therefore we have elegant tuning knob fitted to the geometry of the dot.

\subsubsection{Wave-function}

On the other side, let us focus on the qualitative discussion of the effect of the decreasing velocity on the spatial distribution of the wave function.
Realizing Eq.~\eqref{eq:commutator}, we can argue that lowering twisting angle renormalizes the effective velocity towards lower values and the state becomes better localized and also slightly shifted towards the boundary of the dot.\\
\begin{figure}[htpb]
\centering
\includegraphics[scale = 0.5]{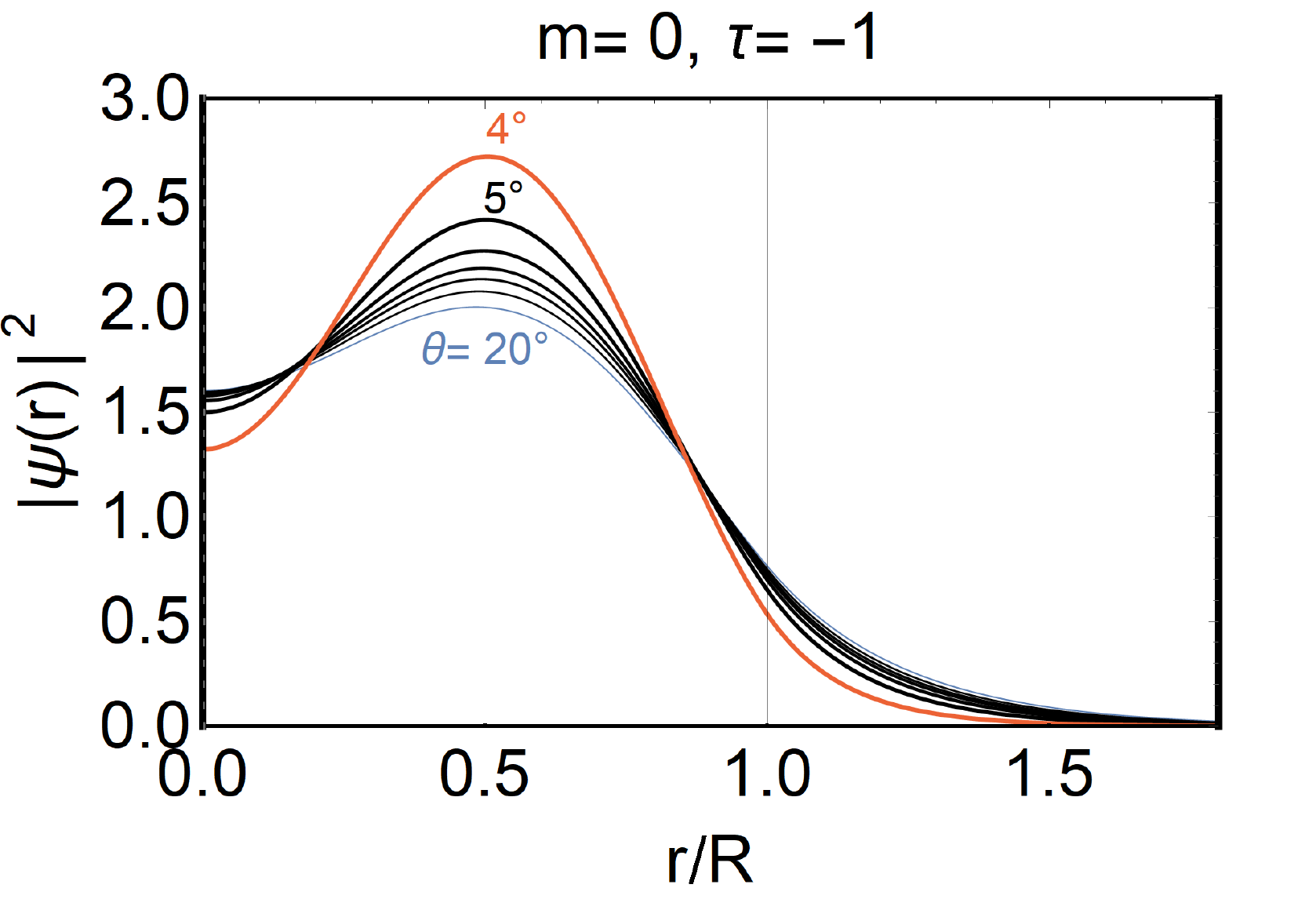}
\caption{Wave function of the twisted bilayer quantum dot as a function of the twisting angle $\theta$. Parameters related to the considered model: $m = 0$, $\tau = 1$, $U_0 = \Delta = 2\delta$, $R = 25\, nm$, $\rho = 1$, and the twisting angle is changing from $\theta = 4^{\circ}$ (top curve) to $\theta = 20^{\circ}$ (bottom curve).}
\label{fig:Wave_fun}
\end{figure}

In the Fig.~\ref{fig:Wave_fun}, we plot the probability distribution in the radial direction, considering finite magnetic field ($B = 2.11\, T$), in order to have better localized state. The value of the magnetic field $B$ is chosen in the way that we have exactly one allowed state in the dot with the radius $R$, since $\rho = 1$. Notice, that slightly increased localizability of the realised state towards the boundary of the dot as a function of the decreasing $\theta$ is also in qualitative agreement with the numerical analysis published in the Ref.~\onlinecite{Tiutiunnyka_2019}.\\

Let us emphasize at the end of this subsection, that our numerically obtained results, focusing on the behavior of the energy states as well as envelope wave-function in the BGM-QD system under the large twisting angles $\theta$ nicely match with the expectations based on the approach discussed in the Subsections.~\ref{subsec:Up} and~\ref{subsec:redEnScale}.

\subsubsection{Comment on the small twisting angles}

From what we know so far, we can qualitatively address the question of the small twisting angles $\theta$, or i.e. situations when $R \approx D$. In the cases, when $R \gg D$ the dot system is fully described either by AA or AB stacking together with their states. In the cases, when $R$ starts to be comparable with $D$, based on our calculations we know that the wave function can be localized in the region close to the boundary of the dot. On the other side, described region also corresponds to the huge mixture of commensurate and incommensurate parts of double layer alignment (partially, we can observe such a situation in the subfigures a) and d) of the Fig.~\ref{fig:Moire_boundary}). Therefore, the solution of the problem will contain signatures of solutions of the exact AA or AB stacked limit cases and also signature of our obtained solution considering limit $R \gg D$.\\

No wonder, that in this intermediate regime, the evolution of the energy levels as a function of $\theta$ will depend on the exact (from the direction of the limit values of the twisting angles $\theta = \{ 0^{\circ}, 60^{\circ}\}$) AA or AB stacked state, as well as on the exact geometry of the twisted bilayer (from the limit cases of the solution realising the BGM-QD). Since the geometry on the boundary will be changing abruptly with small changes in angle~$\theta$, we should expect large changes in the evolution of the energy levels. Notice, that this kind of behavior is visible in the already mentioned tight binding study in Ref.~\onlinecite{Mirzakhani_2020} (however not recognized or highlighted by the original authors). Notice also, that considering the angles with diminishing Fermi velocity, assumption of the linearized dispersion (which we stand on) fails completely.

\section{Summary}

Since the numerical results from the solution of the boundary condition considering by a large angle twisted bilayer graphene quantum (Moire) dot problem were already discussed in the previous section, let us focus on its synthesis aiming for a better understanding of this highly tunable mesoscopic system. The main message of this letter is therefore following: twisting the BGM-QD system in the regime of large angles shows to be suitable tool in order to get better localized states. Such a state also has lower energy due to reduced velocity.\\

Added value of our work is also in the effort to shed more analytical light into the field of twistronics, which is currently dominated by the numerical studies based on the first-principle calculations already discussed in the referenced material\cite{Recher_2009, Mirzakhani_2020}. As we can notice, some of the behavior of the energy levels as well as corresponding states can be examined and guessed by simple analytic analysis provided in the Sec.~\ref{sec:BGM-QD}.\\

Use of the better localized states is natural in the tunneling applications \cite{Wenz_2019, Ubbelohde_2015} (and references therein). Energy levels distance on the infrared scale ranging from $\delta \approx 7\, meV$ (considering $R = 100\, nm$) already up to $\delta \approx 700\, meV$ (considering $R = 1\, nm$) craves for the application in the infrared spectroscopy area. Already discussed energy scale of few $meV$ also opens questions in the direction of exploration of superconductive phenomena on quantum dot systems, since the system itself is (with the considered length scale) able to host a Cooper pair. In this way, it shows that the technology of the gate bounded quantum dots enriches already vast area of the quantum dot research \cite{Holloway_2010}.\\

It is well known, that one of the problems of graphene regarding commercial applications is production of the large sheets of this material. However, small flakes suitable for the quantum dots are not a problem at all. Let us just mention recent development in the field of graphene dot physics towards the identification of the suitable two dimensional dot flakes \cite{Greplova_2020} and let us connect this potential with the progress of the understanding of the measured states \cite{Volk_2011, Eich_2018, Kurzmann_2019, Bucko_2020}. We should also definitely not omit successful experimental effort in the fabrication of the large angle twisted samples motivated by the beauty of the dodecagonal quasicrystal pattern appearance at $\theta = 30^{\circ}$ \cite{Ahn_2018, Pezzini_2020}, and its richness on the Dirac cone replicas structure.\\

All of the mentioned ideas show promising research future in the field of the twisted bilayer graphene quantum dot physics. Our modest contribution to the fast developing field lays in the direction of the very simple description of the tuning by the large angle twisting of two graphene flakes placed above each other. It shows, that this effect leads to the decrease of the electron velocities. Described effect causes several features in the energy and state description including e.g. higher localization of the envelope wave-function.\\

\section*{Acknowledgements}
We are grateful for financial support provided by the Slovak Research and Development Agency under Contract No. APVV-19-0371 and by the agency VEGA under Contract No. 1/0640/20. We are also grateful for financial support from the Swiss National Science Foundation through Division II (Grant No. 184739). We are also grateful to Eliska Greplova, Peter Rickhaus, Fokko de Vries and Frank Schäfer for many useful and interesting comments and discussions.

\begin{appendix}

\section{Solution strategy}\label{sec:Appendix_A_Sol_Strategy}

In this Appendix, we further simplify the set of the differential equations defined by the Eq.~\eqref{eq:H0b_1} using assumption of the polar symmetry of the confining potential $U(\boldsymbol{r})$, as well as further matrix structure of the considered equations. At the end of the Appendix, we also provide full explanation of the boundary condition defined by the Eq.~\eqref{eq:Boundary}.\\

Due to the rotational symmetry of our problem, let us formulate the kinetic part of $H_{\tau}(\theta)$, which applies to $\Psi_1(r)$, using polar coordinates:
\begin{equation}
H_0(\theta) =\\
\frac{\hbar v_F(\theta)}{i \sqrt{2} l_B }
\begin{pmatrix}
0 & \partial_{\xi} - \frac{m-1/2}{\xi} -s\xi \\
\partial_{\xi} + \frac{m-1/2}{\xi} +s\xi & 0
\end{pmatrix},
\end{equation}
where we introduced dimensionless units of $\xi = r/\sqrt{2}l_B$ and magnetic length $l_B = \sqrt{\hbar/e B}$, which is at the value of $B =1\,T$ on the scale of $l_B \approx 26\, nm$.\\

Two first order equations for components of $\Psi_1(r)$ have known solutions in terms of the iterative relations \cite{abramowitz_stegun_2013}:
\begin{eqnarray}\label{reca}
\label{eq:rec1}\left(\partial_{\xi} - (m-1/2)/ \xi -s \xi\right) \phi_{m-1}^s     &=&   a_{1}^{s} \,\phi_{m}^s, \nonumber\\
\label{eq:rec2}\left(\partial_{\xi} +(m-1/2)/ \xi +s \xi\right) \phi_{m}^s  & =&  a_{2}^{s} \,\phi_{m-1}^s.
\end{eqnarray}
The functions $\phi_m^s$ as well as coefficients $a^s_i$ differ inside and outside the dot. $\phi_m^s$ are proportional to confluent hypergeometric (also known as Kummers) functions:
\begin{eqnarray}
 \phi_{m+\alpha}^{s}(\xi) &\equiv& \begin{cases}
                            e^{-\xi^2 /2} \xi^{b-\frac{1}{2}} M(a,b,z) /\Gamma(b), & r\leq R \\[10pt]
                            e^{-\xi^2 /2} \xi^{b-\frac{1}{2}} U(a,b,z), & r > R
                            \end{cases}
\end{eqnarray}
in which we used standard notation for confluent hypergeometric functions of the first $M(a,b,z)$ and second $U(a,b,z)$ kind. In our case:
\begin{eqnarray}
a &=& \frac{|m+\alpha|+1+s (m-1-\alpha) }{2}+\frac{\kappa^2}{4},\nonumber\\
b &=& 1+|m+\alpha|,\nonumber\\ 
z &=& \xi^2.
\end{eqnarray}
As for the coefficients $a_i^s$ defined in the Ref.~\onlinecite{abramowitz_stegun_2013} inside the dot:
\begin{eqnarray}
    a_1^s &= \begin{cases}
                \kappa^2/2, & m \geq 1\\
                2, & m = 0\\
                2, & m \leq -1
            \end{cases}\quad
    a_2^s &= \begin{cases}
                2, & m \geq 1\\
                \kappa^2/2, & m = 0\\
                \kappa^2/2, & m \leq -1
            \end{cases}
\end{eqnarray}
For $r>R$ we can observe that the coefficients are independent from angular momentum $m$:
\begin{eqnarray}
a_{1}^{s}  & =&   -[(s+1)+\kappa^2(1-s)/4], \nonumber \\
a_{2}^{s}  & =&  -[(1-s)+\kappa^2(1+s)/4].
\end{eqnarray}
Next, we can further factorize two-site envelope wave function $\Psi^{\tau}_1$ to introduce $\Psi^{\tau}_2$: 
\begin{equation}
\label{eq:transform2}
\Psi^{\tau}_1 =
\begin{pmatrix}
\phi_{m}^s & 0 \\
0& \phi_{m-1}^s
\end{pmatrix} \Psi^{\tau}_2.
\end{equation}
With this choice and using formulae (\ref{eq:rec1}) -- (\ref{eq:transform2}) we can manipulate the equations so that we can further simplify and replace:
\begin{eqnarray}
H_0 \Psi^{\tau}_1 &=& H_0 \begin{pmatrix}
\phi_{m}^s & 0 \\
0& \phi_{m-1}^s
\end{pmatrix} \Psi^{\tau}_2 \nonumber \\
&=& \begin{pmatrix}
\phi_{m}^s & 0 \\
0& \phi_{m-1}^s
\end{pmatrix}
\frac{\hbar v_{F}(\theta)}{i \sqrt{2} l_B }
\begin{pmatrix}
0 & a_{1}^{s} \\
a_{2}^{s} & 0
\end{pmatrix}
\Psi^{\tau}_2. \nonumber
\end{eqnarray}\\

After all these steps, we end up with the factorization of the original  two-site envelope wave-function: 
\begin{equation}
    \Psi^{\tau}(r,\varphi) = \frac{e^{ i m \varphi}}{\sqrt{r}} 
\begin{pmatrix}
1 & 0 \\
0& e^{- i \varphi}
\end{pmatrix}
\begin{pmatrix}
\phi_{m}^s(r) & 0 \\
0& \phi_{m-1}^s(r)
\end{pmatrix} \Psi^{\tau}_2.
\label{eq:spinor_expansion}
\end{equation}
The radial dependence of  the envelope function is incorporated in $\phi^s_i$ functions. On the other hand, $\Psi^{\tau}_2$ is not coordinate dependent. Moreover, interaction part of the hamiltonian commutes with both matrices in the above envelope function expansion and thus we can move $H_1^\tau$ inside the envelope function in front of $\Psi^{\tau}_2$. Then we are left with the task:
\begin{equation}
    \label{eq:final_schrod}
    \frac{\hbar v_F(\theta)}{i \sqrt{2} l_B }
\begin{pmatrix}
0 & a_{1}^{s} \\
a_{2}^{s} & 0
\end{pmatrix}\Psi^{\tau}_2 + H_1^\tau \Psi^{\tau}_2 = E \Psi^{\tau}_2,
\end{equation}
which is equivalent to solving the following homogeneous system:
\begin{equation}
\begin{pmatrix}
\tau\Delta + U(r) - E &  a_{1}^{s}\frac{\hbar v_F(\theta)}{i\sqrt{2}l_{B}}  \\
a_{2}^{s} \frac{\hbar v_F(\theta)}{i\sqrt{2}l_{B}} & -\tau\Delta + U(r) - E 
\end{pmatrix}
\Psi^{\tau}_2 = 0.
\label{eq:4x4forPsi2}
\end{equation}
We have the condition for zero determinant in (\ref{eq:4x4forPsi2}) in order to have nontrivial $\Psi^{\tau}_2$. Full envelope function can be then easily recovered. However, we have two free parameters to fix, namely $\kappa$ (present in $a^s_i$) and $E$. One can be fixed by the condition on singularity of the above matrix which yields the following relation between $\kappa$ and $E$:
\begin{equation}\label{eq:kappas}
\kappa_{<,>}^{2} = 2 l_B^2\left(\Delta^{2} - \varepsilon_{<,>}^{2}\right)/(\hbar v_F(\theta))^2.
\end{equation}
Throughout the task we set:
\begin{equation}
    U(r) = \begin{cases}
            0, &r \leq R \\
            U_0, &r>R 
            \end{cases}
\end{equation}
and therefore $\varepsilon_< = E$, resp. $\varepsilon_> = E-U_0$. The second condition to fix remaining degree of freedom comes from the continuity of two-site envelope wave-function at the boundary of the dot.\\

In both cases (inside and outside) we have just one value of $\kappa$. In this way we basically have solution inside $\Psi^{\tau}_{1,<}(\xi)$ and outside $\Psi^{\tau}_{1,>}(\xi)$ of the dot up to a multiplicative constants (let us call them $C_{<}$ and $C_{>}$). We find the energy of the respective state by matching these two at $r = R$. Condition on the nontrivial solution with nonzero $C_{<}$ and $C_{>}$ immediately leads to linear combination of the realized envelope wave-function. So, the solution which discloses the allowed values of $R$ (or $B$, or $\theta$) and $E$ can be found by solving for zero of the determinant:
\begin{equation}
    Det^{\tau}(R,B,\theta,E) = \vert \Psi^{\tau}_{1,<}(R,B,\theta,E), \Psi^{\tau}_{1,>}(R,B,\theta,E)\vert.
\end{equation}

\noindent Ratio of the considered constants can be easily found from the ratio of the equations in the boundary condition and the last fix is coming from the normalization condition of the envelope wave-function.

\end{appendix}

\bibliography{References}

\end{document}